\documentclass[%
 aip,
 amsmath,amssymb,
 reprint,%
]{revtex4-1}

\usepackage{graphicx}
\usepackage{dcolumn}
\usepackage{bm}
\usepackage[colorlinks,linkcolor=blue,anchorcolor=blue,citecolor=blue,urlcolor=blue]{hyperref}
\usepackage[utf8]{inputenc}
\usepackage[T1]{fontenc}
\usepackage{mathptmx}

\newcommand{\ket}[1]{|#1\rangle}

\begin{document}

\preprint{AIP/123-QED}

\title{A high-sensitivity fiber-coupled diamond magnetometer with surface coating}

\author{Shao-Chun Zhang}
\author{Hao-Bin Lin}
\author{Yang Dong}
\author{Bo Du}
\affiliation{{CAS Key Laboratory of Quantum Information, University of Science and Technology of China, Hefei,
230026, P.R. China}}
\affiliation{{CAS Center For Excellence in Quantum Information and Quantum Physics, University of Science
and Technology of China, Hefei, 230026, P.R. China}}
\author{Xue-Dong Gao}
\author{Cui Yu}
\author{Zhi-Hong Feng}
\affiliation{{National Key Laboratory of ASIC, Hebei Semiconductor Research Institute, Shijiazhuang 050051, P.R. China}}
\author{Xiang-Dong Chen}
\author{Guang-Can Guo}
\author{Fang-Wen Sun}
\email{fwsun@ustc.edu.cn}
\affiliation{{CAS Key Laboratory of Quantum Information, University of Science and Technology of China, Hefei,
230026, P.R. China}}
\affiliation{{CAS Center For Excellence in Quantum Information and Quantum Physics, University of Science
and Technology of China, Hefei, 230026, P.R. China}}

\date{\today}

\begin{abstract}
Nitrogen-vacancy quantum defects in diamond offer a promising platform for magnetometry because of their remarkable optical and spin properties.
In this Letter, we present a high-sensitivity and wide-bandwidth fiber-based quantum magnetometer for practical applications. By coating the diamond surface with silver reflective film, both the fluorescence collection and excitation efficiency are enhanced. Additionally, tracking pulsed optically detected magnetic resonance spectrum allowed a magnetic field sensitivity of $35$ pT$/\sqrt{\rm{Hz}}$ and a bandwidth of $4.1$ KHz. Finally, this magnetometer was successfully applied to map the magnetic field induced by the current-carrying copper-wire mesh. Such a stable and compact magnetometry can provide a powerful tool in many areas of physical, chemical, and biological researches.
\end{abstract}

\maketitle

Quantum sensors based on nitrogen-vacancy (NV) centers in diamond offer a powerful platform for magnetometry across a range of length scale \cite{barry2020sensitivity}. At the nanometer scale, benefiting from their angstrom-scale size, single NV centers have been used for applications with ultra-high-resolution \cite{taylor2008highsensitivity,lesage2013optical,mcguinness2011quantum,abobeih2019,balasubramanian2008nanoscale,degen2008scanning,chen2019superresolution,chen2015subdiffraction} in living  biology systems \cite{lesage2013optical, mcguinness2011quantum}, nuclear magnetic resonance \cite{abobeih2019}, and magnetic resonance force microscopy \cite{balasubramanian2008nanoscale,degen2008scanning}. At the micrometer scale, sensor employing ensemble of NV centers usually provide improved sensitivity at the cost of spatial resolution, which facilitates a wide-field magnetic imaging for bio-magnetic structures \cite{barry2016optical}, integrated circuits \cite{turner2020magnetic}, and even the geological samples \cite{glenn2017micrometerscale}. At the millimeter scale, bulk ensemble magnetometers with large active volumes offer ultra-high sensitivity \cite{chatzidrosos2017miniature,jensen2014cavityenhanced, wolf2015subpicotesla}, but most of the techniques have the requirement of cavity \cite{chatzidrosos2017miniature,jensen2014cavityenhanced} or parabolic lens \cite{wolf2015subpicotesla} for increasing fluorescence collection efficiency, leading the sensors unsuitable for using at the few-millimeter scale.

The recently developed fiber-optic probes \cite{fedotov2014fiberoptic, fedotov2014fiberoptic, blakley2016fiberoptic, blakley2016fiberoptic, patel2020subnanotesla, duan2019efficient,duan2018enhancing, fedotov2020alloptical,zhang2019thermaldemagnetizationenhanced,zhang2020robust} coupled with NV centers enabled a compact approach at the scale of a few hundred micrometers. Coupling the NV center to an optical fiber integrated with a two-wire microwave transmission line enable a sensitivity of $300$ nT$/\sqrt{\rm{Hz}}$ \cite{fedotov2014fiberoptic}. The subsequent double-fiber coupling approach allows efficient noise cancellation and a sensitivity of $35$ nT$/\sqrt{\rm{Hz}}$ \cite{blakley2016fiberoptic}. Additionally, by optimizing fiber collection system, the best sensitivity of $310$ pT$/\sqrt{\rm{Hz}}$ in the frequency range of $10-150$ Hz was achieved \cite{patel2020subnanotesla}. On the other hand, recent study with a matched micro-concave mirror to a sphered optic-fiber end has achieved over $25$ times more fluorescence collection from NV enriched micrometer-sized diamond \cite{duan2019efficient,duan2018enhancing}. In particular, thanks to the simplicity and robustness, such a fiber-optic quantum sensor has successfully realized in vivo detection \cite{fedotov2020alloptical}.

\begin{figure}[htb]
\includegraphics[width=0.45\textwidth]{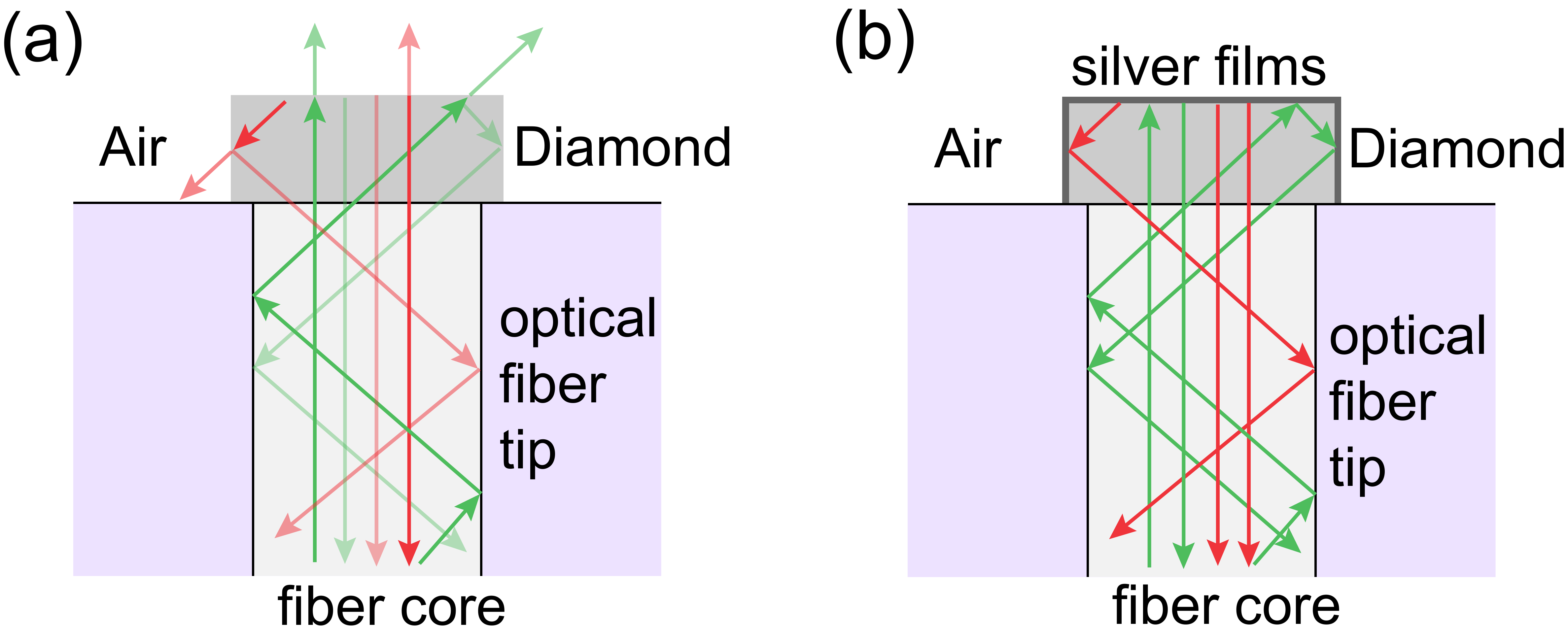}
\caption{\label{SilverFilm}(a-b) Schematic of the fluorescence collection and excitation efficiency with (a) and without (b) the silver films coated on the diamond surfaces.}
\end{figure}

Here, we pasted a bulk diamond on the optical fiber tip, and a reflective film was coated on the five surfaces exposed to the air through silver mirror reaction, as shown in Fig.~\ref{SilverFilm}, which suppressed the green laser and red fluorescence from refracting into the air. Such a strategy can not only enhance the fluorescence collection efficiency, but also the excitation efficiency of NVs in diamond. In this case, we present a fiber-based quantum magnetometer for practical applications with high sensitivity and wide bandwidth. With pulsed optically detected magnetic resonance (ODMR) implementation, optical and microwave (MW) power broadening of the spin resonances can be avoided, which allows a further improvement in sensitivity. By optimizing the initialization time of NV center and then tracking pulsed ODMR, a magnetic field sensitivity of $35$ pT$/\sqrt{\rm{Hz}}$ and a bandwidth of $4.1$ KHz were achieved. Finally, this magnetometer was successfully applied to map the magnetic field induced by the current-carrying copper-wire mesh. Compared with previous measurements based on the lock-in ODMR \cite{fedotov2014fiberoptic, fedotov2014fiberoptic, blakley2016fiberoptic, blakley2016fiberoptic, patel2020subnanotesla, duan2019efficient,duan2018enhancing, fedotov2020alloptical,zhang2019thermaldemagnetizationenhanced,zhang2020robust}, our work has higher sensitivity, wider bandwidth and more ways to optimize the sensitivity, like double quantum magnetometry and spin-bath driving \cite{bauch2018ultralong}. These results make a significant advance in transiting lab-based systems to practical applications at the millimeter scale.

\begin{figure*}
\includegraphics[width=0.8\textwidth]{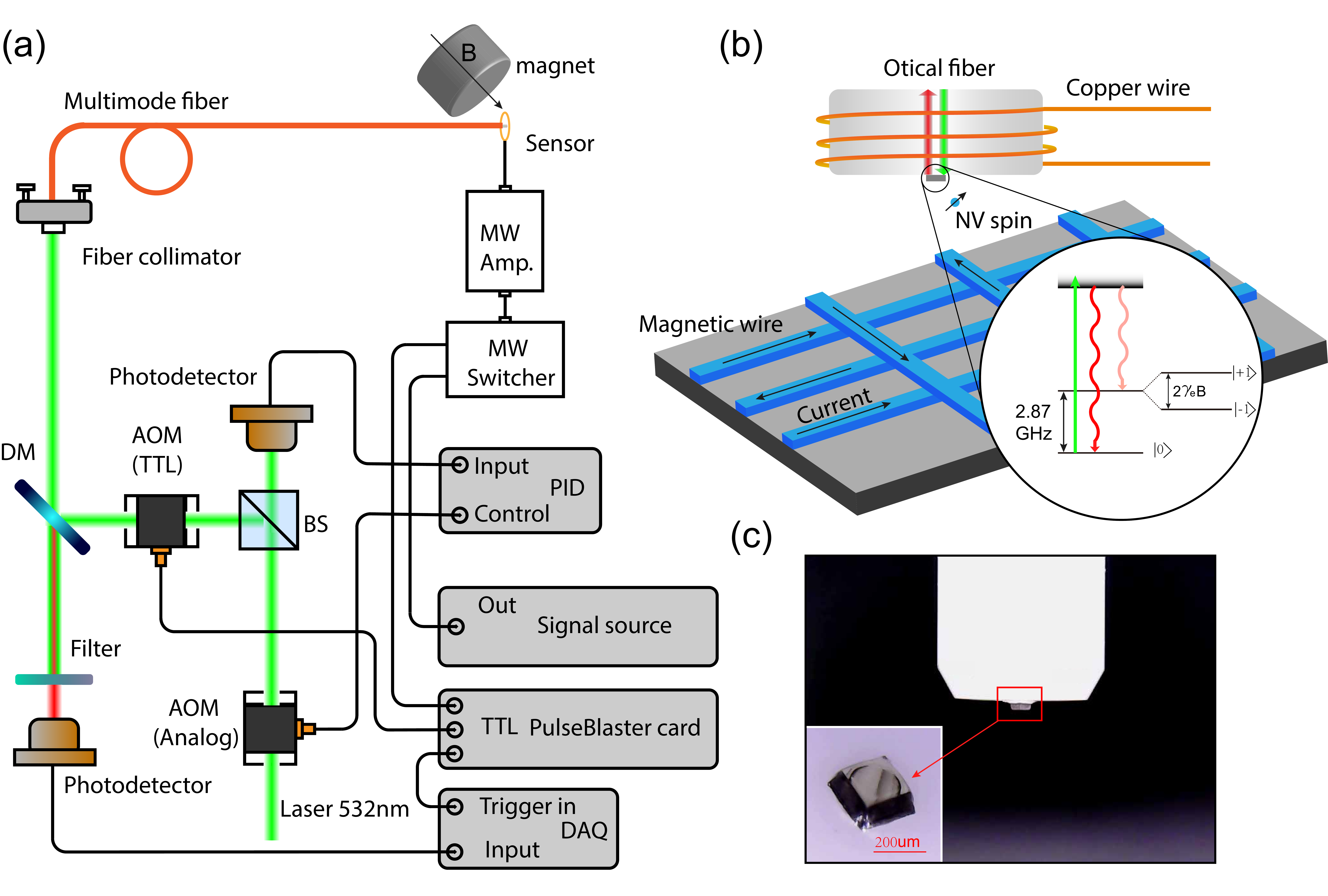}
\caption{\label{experimentsetup}(a) The schematic of fiber-optical magnetometer setup.
DM, long-pass dichroic mirror with the edge wavelength of $658.8$ nm.
(b) Simplified schematic of applications with fiber-optical magnetometer. Magnetic field measurements are performed by a bulk of diamond with NV ensemble placed at the apex of a fiber tip. Current-carrying copper-wire mesh on the glass surface are connected in series with a power supply.  (Inset) Optical excitation of the NV center into spin state $\ket{0}$. The fluorescence starting from state $\ket{0}$ is stronger than that of states $\ket{\pm1}$, allowing the NV spin to be read out optically.  (c) A bulk diamond is attached on the tip of a multi-mode optical fiber. The inset shows the bulk diamond with the size of $200 \times$200$\times$100 $\mu$m$^3$ under the microscope without silver film coated.}
\end{figure*}


Conventional confocal scanning system employed in a sensing scheme poses complications in terms of collective control, signal readout and even the stability, leading to the lab-based demonstrations. Here, these issues can be addressed by using a homebuilt fiber system to excite and detect the NV centers, as shown in Fig.~\ref{experimentsetup}(a). A diamond attached on the tip of a multi-mode optical fiber with a core diameter of $200$ $\mu$m has been mechanically polished and cut into a membrane with dimensions of $200 \times$200$\times$100 $\mu$m$^3$, as shown in Fig.~\ref{experimentsetup}(c). The NV center ensemble in the diamond consisted of [N] $\approx$ $55$ ppm and [NV$^-$] $\approx$ $2$ ppm with [100] surface orientation was grown by plasma assisted chemical vapor deposition. At room temperature, NV$^-$ exhibits an efficient and photostable red PL, which enables optical detection. The ground state is a spin triplet including a singlet state $\ket{0}$ and a doublet state $\ket{\pm{1}}$ separated by a temperature-dependent zero-field splitting (ZFS) $D=2.87$ GHz in the absence of magnetic field, as shown in the inset of Fig.~\ref{experimentsetup}(b). Applying a static magnetic field along the NV axis leads to a splitting of $\ket{-1}$ and $\ket{+1}$ states.

 In the experiment, $532$ nm green laser was sent through an acousto-optic modulator (AOM, AA optoelectronic MT250-A0.5-VIS, Analog modulation) and finally coupled into the multi-mode optical fiber. The power was adjusted with a beam-splitter (BS) cube before the diamond, where part of the laser light was split-off and measured with a photodetector (PD, Thorlabs PDA36A). The signal was then input into a proportional-integral-derivative controller (PID, SRS SIM960) to stabilize the laser power. For the sensing protocol described here, laser pulses on the microsecond timescale for initializing and reading out the NVs are realized by another AOM (AA optoelectronic MT250-A0.5-VIS, TTL modulation). Collected by the same fiber, the photoluminescence (PL) passed through a $647$ nm long pass filter and was finally sent to another photodetector (Thorlabs APD430A). Moreover, the output MW was sent through a switch (M-C ZASWA-2-50DR+) to a high-power amplifier (M-C ZHL-16W-43), and finally delivered by a five-turn copper loop with an outer diameter of $0.5$ mm wound around the optical fiber ceramic plug core, as shown in Fig.~\ref{experimentsetup}(b). All pulse sequences are generated by a pulse blaster card, which are used to both control the timing of the data acquisition (DAQ, NI USB 6343) unit in the experiment and switch on and off the laser and MW sources. 

\begin{figure}[t]
\includegraphics[width=0.45\textwidth]{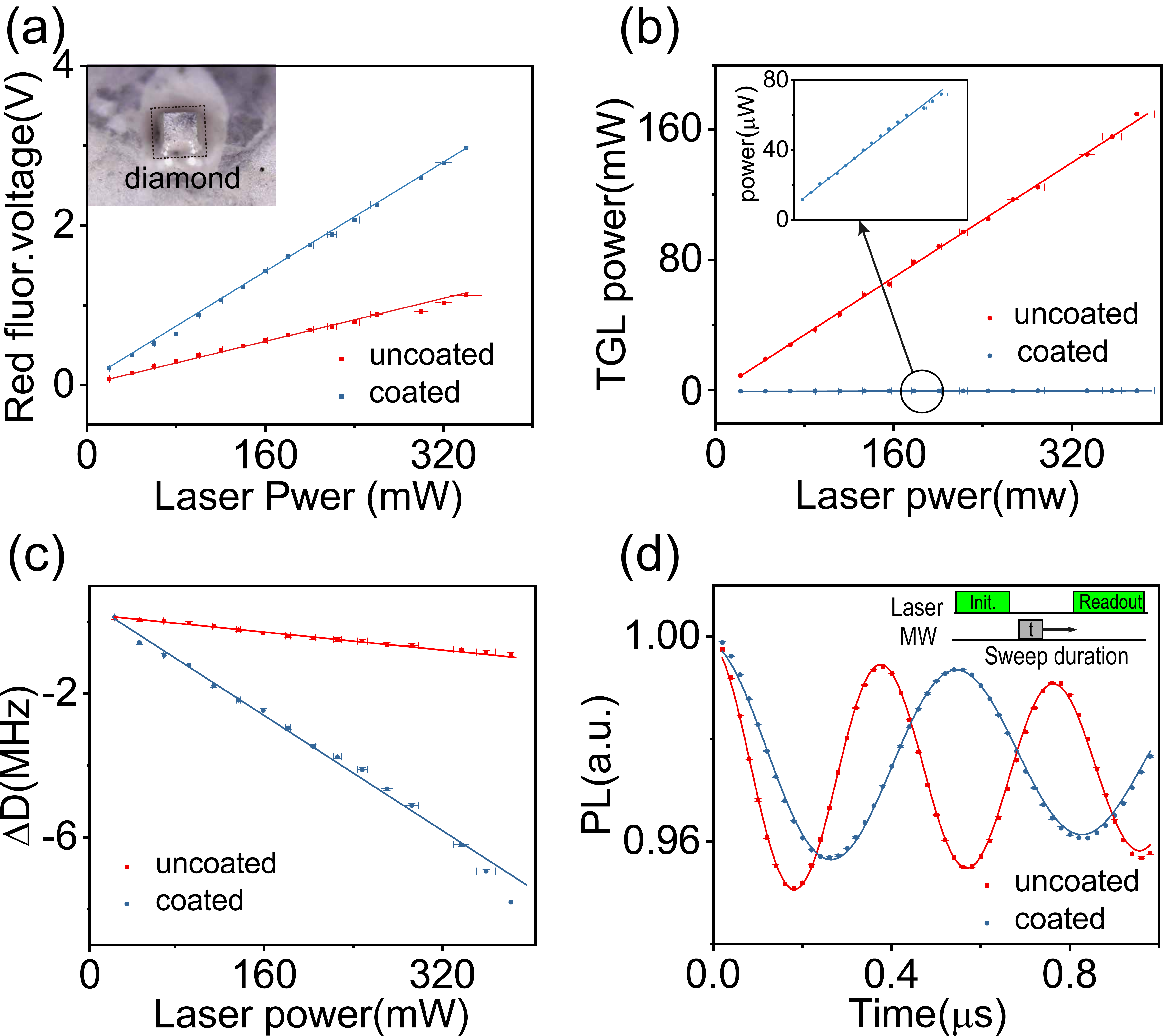}
\caption{\label{Fluorescence} (a) Red fluorescence collection with and without the reflective coating, respectively. The inset shows the diamond coated with silver on the tip of fiber, and the fluorescence collection efficiency can be enhanced by $2.5$ times. (b) The power of transmitted green light (TGL) from the diamond. The inset shows the green light power of coated fiber with the unit of $\mu$W. (c) The ZFS changes $\Delta$D as a function of the pump green laser power. (d) Rabi oscillations with the pump laser power of $320$ mW. The inset show the pulse sequence. The silver film on the diamond surface has affected the microwave radiation.}
\end{figure}

The reflective film coated on the five surfaces exposed to the air can be easily and quickly realized by silver mirror reaction, as shown in the inset of Fig.~\ref{Fluorescence}(a). Moreover, Fig.~\ref{Fluorescence}(a) shows the red fluorescence collection as a function of green pump laser power. Far from saturation, fluorescence increases linearly with the laser power. The ratio of the slopes of the two lines represents a $2.5$-fold increase in the collected fluorescence. On the other hand, almost half of the green light was transmitted from the diamond into the air in the absence of reflective film, leading to a low fluorescence excitation efficiency, as shown in Fig.~\ref{Fluorescence}(b). After coating treatment, almost no green light was transmitted, which indicates the perfect reflection effect of silver film. In general, diamonds containing high concentrations of nitrogen are dark black, resulting in the diamond absorbing part of the green pump laser and temperature rising. The red dots in Fig.~\ref{Fluorescence}(c) shows the linear change of ZFS with laser power, which is consistent with the results of previous studies \cite{zhang2019thermaldemagnetizationenhanced}. Note that $dD/dT\approx-74\rm{KHz/K}$ at room temperature \cite{chen2011temperature,acosta2010temperature}. Although coating reflective films on the diamond could double the green light absorbed by the diamond, the green light heated the silver films and thus led to temperature change slope more than twice that of the uncoated diamond [blue dots in Fig.~\ref{Fluorescence}(c)]. In order to determine the impact of silver films on the MW field, a $392$ G magnetic field was applied along one of the NV axes. And Rabi oscillations were then performed under the same MW input power, as shown in Fig.~\ref{Fluorescence}(d). The inset shows the Rabi pulse sequence with the initialization time of $500$ $\mu$s, AOM polarization time of $1.5$ $\mu$s and DAQ unit readout pulses length of $300$ ns. In the Rabi oscillation experiment, the MW frequency was tuned to match the NV spin resonance (from $m_s = 0$ to $m_s = -1$ transition) and the NV fluorescence was measured as a function of the MW pulse duration. The reduced Rabi frequency indicates that the silver film on the diamond surface attenuates the MW field, but it does not affect the subsequent sensing performance.

\begin{figure}[b]
\includegraphics[width=0.43\textwidth]{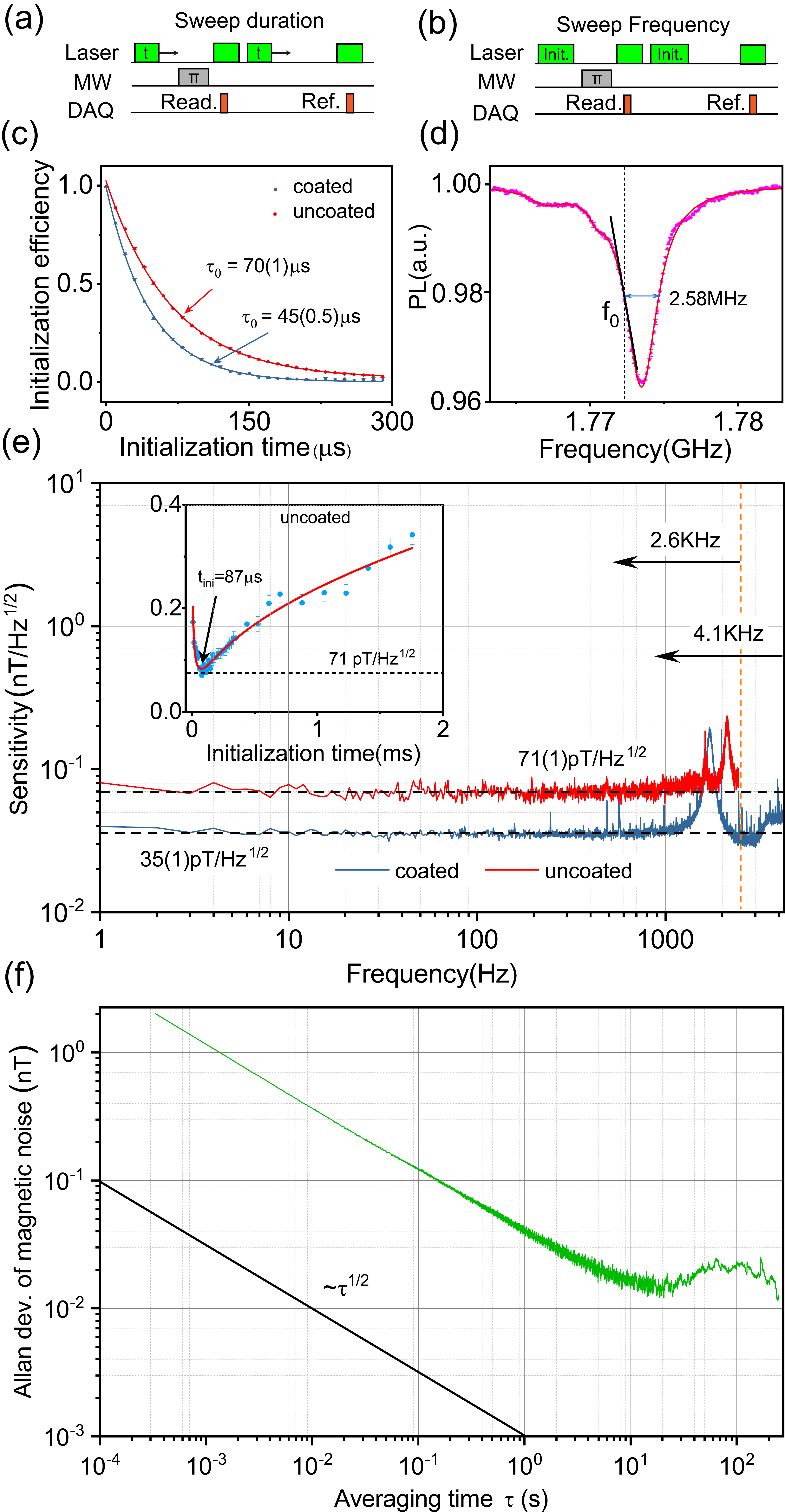}
\caption{\label{Sensitivity} (a) Pulse sequences of initialization time sweeping. (b) Pulse sequences of pulsed ODMR. (c) The initialization efficiency as a function of laser duration time $t$. Both of the two curves are fitted by exponential decay function. (d) The pulsed ODMR with FWHM of $2.58$ MHz, where $f_0$ is at the point of maximum slope. (e) Magnetic field sensitivities of coated and uncoated fiber-optical magnetometer. The sensitivity is optimized from $71(1)$ pT$/\sqrt{\rm{Hz}}$ to $35(1)$ pT$/\sqrt{\rm{Hz}}$, with the bandwidth is broadened from $2.6$ KHz to $4.1$ KHz. The inset shows the magnetic field sensitivity of uncoated fiber-optical magnetometer as a function of initialization time $t_{\rm{ini}}$ and was fitted by Eq.~(\ref{SensitivityEquation2}), where $t_{\rm{ini}}\approx1.25\tau_0$ gives the best sensitivity. (f) Scaling of Allan deviation from the pulsed ODMR sequence with the reflective coating.}
\end{figure}

Coating reflective films on the diamond surface exposed to the air can not only increase the fluorescence collection efficiency, but also excitation. The latter can be proved by measuring the initialization time of the NV center \cite{barry2020sensitivity,ahmadi2018nitrogenvacancy}. The measurement pulse sequence is depicted schematically in Fig.~\ref{Sensitivity}(a). The NV$^-$ spin state is first optically initialized to $m_{s}=0$, and the initialization efficiency is laser power and pulse length dependent \cite{ahmadi2018nitrogenvacancy}. For common noise cancellation, the pulse sequence is applied twice except the MW off in the second pulse \cite{bucher2019quantum}. In this case, the division of two signals denotes the result of the detection sequence executed once. Setting the pump laser power to $320$ mW and sweeping the initialized duration, the initialization efficiency or the measurement contrast increases as $1-e^{-t/\tau_0}$, as illustrated in Fig.~\ref{Sensitivity}(b), where $\tau_{0, \rm{uncoated}}$ and $\tau_{0, \rm{coated}}$ are $71(1)$ $\mu$s and $45(0.5)$ $\mu$s, respectively. Compared with the uncoated diamond, the coated reduced the time for polarizing NV$^-$ spin state. Alternatively, the coated diamond provides a higher contrast with the same pump laser power, and further improves the sensitivity.

In order to measure the magnetic field sensitivity of the fiber-based quantum magnetometer, pulsed ODMR was employed. In contrast to cw ODMR, this technique avoids optical and MW power broadening of spin resonances, enabling nearly $T_2^*$-limited measurements without requiring high Rabi frequency \cite{barry2020sensitivity}. And pulsed ODMR is linearly sensitive to magnetic field variations, making the method attractive when high MW field strengths are not available \cite{dreau2011avoiding}. The pulsed ODMR protocol is depicted schematically in Fig.~\ref{Sensitivity}(c), and the result with a Lorentzian resonance line shape is plotted in Fig.~\ref{Sensitivity}(d). The sensitivity $\eta_B$ of the measurement is given by the following relation \cite{barry2020sensitivity}:
\begin{align}\label{SensitivityEquation1}
\eta_{B} =\frac{4}{3\sqrt{3}}\frac{h}{g_{e}\mu_{B}}\frac{\Delta\nu\sqrt{t_m}}{C\sqrt{R_m}}\text{,}
\end{align}
where $\Delta\nu$ is the full-width at half-maximum (FWHM) of the resonance spectrum, $C$ is the contrast of pulsed ODMR, $t_m$ and $R_m$ correspond to the measurement time and fluorescence photon emission per sequence operation.  Due to the interrogation time and readout pulse are much smaller than the initialization time \cite{wolf2015subpicotesla}, i.e., $t_{\rm{\pi}}+t_{\rm{read}}\ll t_{\rm{init}}$, the measurement duration $t_m\approx 2t_{\rm{init}}$. Additionally, the ODMR contrast is strongly dependent on the initialization time, yielding
\begin{align}\label{SensitivityEquation2}
\eta_{B} \approx \frac{4}{3\sqrt{3}}\frac{h}{g_{e}\mu_{B}}\frac{\Delta\nu}{C_0(1-e^{-t_{\rm{init}}/\tau_0})}\frac{\sqrt{2t_{\rm{init}}}}{\sqrt{R_m}}\text{.}
\end{align}

Hereafter, we assume $R_m$ is independent of $t_{\rm{init}}$ and the choice of $t_{\pi}\approx T_2^*$ ($\approx 0.4\mu s$ here) allows nearly $T_2^*$-limited linewidth while preserving PL contrast\cite{dreau2011avoiding}. Eq.~(\ref{SensitivityEquation2}) also illustrates the benefits attained by optimized $t_{\rm{init}}\approx 1.25\tau_{0}$.
By fixing the MW frequency at point of maximum slop ($f_0$ in Fig.~\ref{Sensitivity}(d)), the output of each pulse sequence operation was recorded for $100$ s. The calculated power spectral density for coated and uncoated diamond are depicted in Fig.~\ref{Sensitivity}(e). For frequencies in the range of $1 \rm{Hz}-2.6 \rm{KHz}$, we reach a magnetic field sensitivity of $71(1)$ pT$/\sqrt{\rm{Hz}}$ with the diamond uncoated. After coating reflective films, the sensitivity was enhanced to $35(1)$ pT$/\sqrt{\rm{Hz}}$ accompanying the  bandwidth broaden to $1$ $\rm{Hz}-4.1$ $\rm{KHz}$. Furthermore, the inset in Fig.~\ref{Sensitivity}(e) shows the sensitivity as a function of initialization time, which is well fitted by Eq.~(\ref{SensitivityEquation2}) with $\tau_{0}=\tau_{0, \rm{uncoated}}$. The Allan deviation from the pulsed ODMR sequence with the reflective coating is plotted in Fig.~\ref{Sensitivity}(f), and the trace exhibits a constant $\tau^{-1/2}$ scaling, which signifies the dominance of stochastic white noise, like thermally induced electronic noise generated in the detector. A minimum floor for this detection is reached for $\tau=10-30$ s, with further averaging (for long detection time $\tau$) giving no advantage mainly due to temperature-based magnetization fluctuations.

\begin{figure}[bt]
\includegraphics[width=0.45\textwidth]{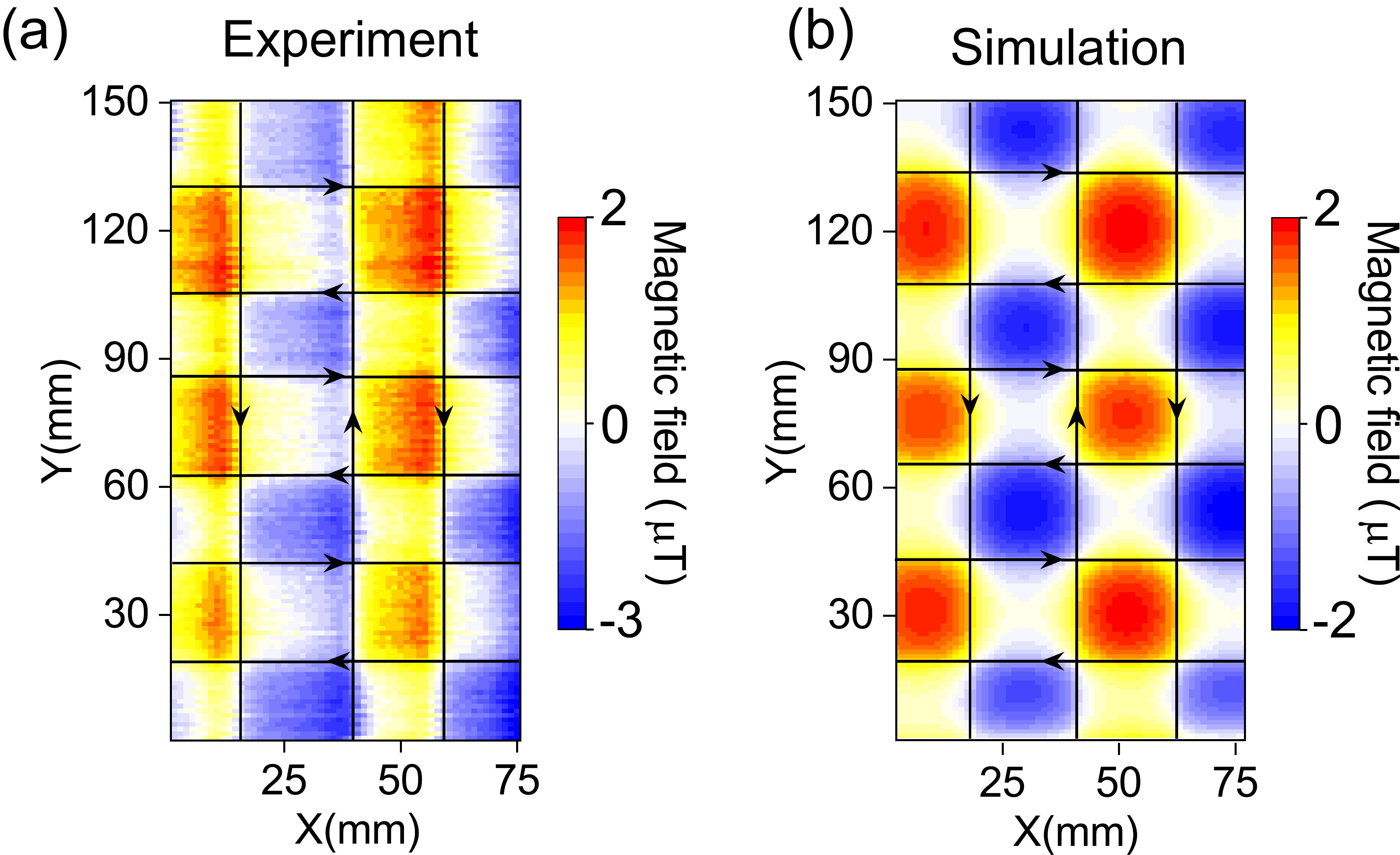}
\caption{\label{MagneticFieldMapping}(a) Mapping the magnetic field induced by the current-carrying copper-wire mesh with coated fiber-optical magnetometer respectively. Arrows indicate the directions of the $1$ mA current. (b) Simulation of the magnetic field in (a) via Biot-Savart-Laplace equation.}
\end{figure}

Finally, the coated magnetometer probe is incorporated into a three-axis motorized translation stage with $1$ $\mu$m accuracy on each axis. In order to verify the feasibility of scanning imaging in the micron size, this magnetometer was applied to map the magnetic field induced by a current-carrying copper-wire mesh with a grid pitch and wire diameter of approximately $2$ mm and $30$ $\mu$m, respectively, as shown in Fig.~\ref{experimentsetup}(b). Before imaging, the probe was placed at a $50$ $\mu$m distance from the mesh, and then, MW frequency was fixed at $f_0$ in Fig.~\ref{Sensitivity}(b), so that the probe had the maximum response to magnetic field variations. Fig.~\ref{MagneticFieldMapping}(a) shows the 2D magnetic field projected onto NV axis under an injected current $1$ mA, where the arrows indicate current directions. Moreover, the magnetic field profile predicted by the Biot-Savart-Laplace equation agrees very well with the experimental measurement, as shown in Fig.~\ref{MagneticFieldMapping}(b).

In conclusion, we have experimentally demonstrated a fiber-based quantum magnetometer coupled with NV centers, which enables practical applications at the micron scale with picotesla sensitivity. Compared with the previous cw ODMR measurement, our magnetometer avoided optical and MW power broadening of the spin resonances in the pulsed ODMR implementation, thus allowing a further improvement in sensitivity. Additionally, coating the diamond surfaces exposed to the air with reflective films benefits sensitivity in two ways: first, the reflective films enhances the fluorescence collection efficiency and thus improve the sensitivity by $\sqrt{\eta_{\rm{eff}}}$\cite{yu2020enhanced}, where $\eta_{\rm{eff}}$ is the efficiency enhancement. And second, the enhanced excitation efficiency of NV center shortens the initialization time, which decreases the fraction of time devoted to spin precession. All of these boost the sensitivity and bandwidth to $35(1)$ pT$/\sqrt{Hz}$ and $4.1$ KHz respectively. Higher laser power can not only enhance fluorescence intensity, but also shorten the initialization time of NV spin. However, the accompanying temperature rise might damage the sample, thus practical considerations may prevent this approach. In addition to pulsed ODMR, Ramsey is an alternative magnetometry method for sensing. With such a protocol, NV spin ensemble $T_2^*$ can be greatly extended by a combination of double quantum magnetometry and spin-bath driving \cite{bauch2018ultralong}, allowing a sub-picotesla sensitivity of dc magnetic-field. In this case, our fiber-based quantum magnetometer can be used to measure biomagnetic signals, e.g., magnetic signals by the action potentials with single-neuron \cite{barry2016optical}, and even the human heart and brain activities \cite{zhang2020recording}.

This work is supported by the National Key Research and Development Program of China (Grant No. 2017YFA0304504), the Science Challenge Project (Grant No. TZ2018003), the National Natural Science Foundation of China (Grants No. 91850102 and No. 12005218), the Anhui Initiative in Quantum Information Technologies (Grant No. AHY130000), and the Fundamental Research Funds for the Central Universities (No. WK2030000020).


%

\end{document}